# A Computational Model of Levodopa-Induced Toxicity in *Substantia Nigra Pars Compacta* in Parkinson's Disease

Vignayanandam Ravindernath Muddapu[1], Karthik Vijayakumar[2], Keerthiga Ramakrishnan[3], V Srinivasa Chakravarthy[1, #]

[1] Department of Biotechnology, Bhupat and Jyothi Mehta School of Biosciences, Indian Institute of Technology Madras, Chennai 600036, Tamil Nadu, India.

[2] Department of Biotechnology, Rajalakshmi Engineering College, Chennai 602105, Tamil Nadu, India.

[3] Division of Psychiatry and Applied Psychology, School of Medicine, University of Nottingham, Nottingham NG8 1BB, United Kingdom.

# Corresponding Author:

Prof V. Srinivasa Chakravarthy, Ph.D.

Room 505, Block 1, Department of Biotechnology,

Bhupat and Mehta Jyothi School of Biosciences,

Indian Institute of Technology Madras, Adyar,

Chennai 600036, Tamil Nadu, India.

Tel: +91 44 2257 4115

Fax: +91 44 2257 4102

Email: schakra@ee.iitm.ac.in

Running Title: Levodopa-Induced Toxicity in SNc

# A Computational Model of Levodopa-Induced Toxicity in *Substantia Nigra Pars Compacta* in Parkinson's Disease

**ABSTRACT**

*Background:* Parkinson's disease (PD) is caused by the progressive loss of dopaminergic cells in substantia nigra pars compacta (SNc). The root cause of this cell loss in PD is still not decisively elucidated. A recent line of thinking traces the cause of PD neurodegeneration to metabolic deficiency. Due to exceptionally high energy demand, SNc neurons exhibit a higher basal metabolic rate and higher oxygen consumption rate, which results in oxidative stress. Recently, we have suggested that the excitotoxic loss of SNc cells might be due to energy deficiency occurring at different levels of neural hierarchy. Levodopa (LDOPA), a precursor of dopamine, which is used as a symptom-relieving treatment for PD, leads to outcomes that are both positive and negative. Several researchers suggested that LDOPA might be harmful to SNc cells due to oxidative stress. The role of LDOPA in the course of PD pathogenesis is still debatable.

*New Method:* We hypothesize that energy deficiency can lead to LDOPA-induced toxicity (LIT) in two ways: by promoting dopamine-induced oxidative stress and by exacerbating excitotoxicity in SNc. We present a multiscale computational model of SNc-striatum system, which will help us in understanding the mechanism behind neurodegeneration postulated above and provides insights for developing disease-modifying therapeutics.

*Results:* It was observed that SNc terminals are more vulnerable to energy deficiency than SNc somas. During LDOPA therapy, it was observed that higher LDOPA dosage results in increased loss of somas and terminals in SNc. It was also observed that co-administration of LDOPA and glutathione (antioxidant) evades LDOPA-induced toxicity in SNc neurons.

*Comparison with Existing Methods:* Our proposed multiscale model of SNc-striatum system is first of its kind, where SNc neuron was modelled at biophysical level, and striatal neurons were modelled at spiking level.

*Conclusions:* We show that our proposed model was able to capture LDOPA-induced toxicity in SNc, caused by energy deficiency.

# 1. INTRODUCTION

Almost all neurodegenerative diseases have a characteristic loss of a certain special type of cells that are vulnerable to death due to metabolic deficiency (Fu et al., 2018; Muddapu et al., 2020). Parkinson's disease (PD) is characterized by loss of dopaminergic neurons in the substantia nigra pars compacta (SNc), which results in cardinal symptoms such as tremor, rigidity, bradykinesia and postural instability (Goldman and Postuma, 2014). The root cause of SNc cell loss in PD is still not decisively elucidated. Recently, a modeling study has been proposed where PD is described to be resulting from the metabolic deficiency in SNc (Muddapu et al., 2019). The vulnerable cells of SNc are projection neurons with large axonal arbors of complex morphologies, requiring a huge amount of energy to maintain information processing activities (Bolam and Pissadaki, 2012; Giguère et al., 2019; Muddapu et al., 2020; Pissadaki and Bolam, 2013). Due to huge energy requirements, SNc neurons exhibit higher basal metabolic rates and higher oxygen consumption rate, which result in oxidative stress (Pacelli et al., 2015). With the help of a computational model, Muddapu et al. (Muddapu et al., 2019) have recently suggested that the excitotoxic loss of SNc cells might be due to energy deficiency occurring at different levels of neural hierarchy – systems, cellular and subcellular.

Levodopa (LDOPA), a precursor of dopamine (DA), is used as a symptom-relieving treatment for PD (Poewe et al., 2010). The usage of LDOPA for PD is still debated due to its side-effects with long-term treatment (Fahn, 2005; Lipski et al., 2011; Thanvi and Lo, 2004). Several researchers suggested that LDOPA might be harmful to SNc cells by a mechanism that probably involves oxidative stress (Carvey et al., 1997; Pardo et al., 1995; Takashima et al., 1999). However, several others proposed that LDOPA might not accentuate neurodegeneration of SNc neurons (Billings et al., 2019; Fahn, 2005; Fahn et al., 2004; Jenner and Brin, 1998) and sometimes acts a neuroprotective agent (Fahn, 2005; Schapira, 2008; Shimozawa et al., 2019) or promote recovery of dopaminergic markers in the striatum (Murer et al., 1998, 1999). After several studies, it is still not clear why LDOPA is not toxic in case of nonparkinsonian human subjects and healthy animals and toxic in PD models of rodents (Agid, 1998; Fahn, 1997; Lipski et al., 2011; Müller et al., 2004; Olanow and Obeso, 2011; Paoletti et al., 2019; Weiner, 2006; Ziv et al., 1997). The beneficial or toxic effects of LDOPA needs to be investigated with more thorough experiments performed at preclinical and clinical levels.

In this paper, we investigate, using computational modelling, the hypothesis that LDOPA-induced toxicity can occur in two ways: by promoting DA-induced oxidative stress (autoxidation-relevant) (Borah and Mohanakumar, 2010; Carvey et al., 1997; Melamed et al., 1998; Pardo et al., 1995; Takashima et al., 1999; Walkinshaw and Waters, 1995) or by exacerbating excitotoxicity in SNc (autoxidation-irrelevant) (Blomeley et al., 2009; Blomeley and Bracci, 2008; Cheng et al., 1996; Pardo et al., 1993; Thornton and Vink, 2015), or by both the mechanisms which might be actually precipitated by energy deficiency. In order to investigate our hypothesis, we propose a multiscale computational model of L-DOPA-induced toxicity in SNc, which will help us in understanding the mechanism behind neurodegeneration due to LDOPA and gives insights for developing disease-modifying therapeutics.

## 2. MATERIALS AND METHODS

The proposed model of levodopa-induced toxicity (LIT) consists of the cortico-basal ganglia system. We model a part of the basal ganglia system comprising the following nuclei: SNc, Striatum, Subthalamic nucleus (STN), and Globus Pallidus externa (GPe). Within the SNc, we separately consider SNc soma (cell body) and SNc terminals (boutons) that make contact with striatal neurons. Within the striatum, we model D1-type receptor-expressing medium spiny neurons of two subtypes: 1) D1-MSNs that release GABA only (D1-MSN) (GABAergic only) and 2) D1-MSNs that release GABA and substance P, D1-MSN (GABAergic (G), and substance P (S)). In the cortex (CTX), the pyramidal neurons are modeled. Neurons in each nucleus are arranged as a two-dimensional lattice (Figure 1). All the simulations were performed by numerical integration using MATLAB (RRID: SCR_001622) with a time step of $0.1 \, ms$.

To implement energy deficiency conditions in the proposed model, the glucose and oxygen inputs were reduced to SNc cells. As the number of SNc neurons under energy deficiency increases, the dopaminergic tone to striatum decreases due to SNc terminal loss. The dopamine deficiency leads to lesser excitation of MSN neurons by pyramidal neurons in the cortex; as a result, SNc neurons get disinhibited. Disinhibition from MSN leads to overactivity of SNc neurons, which in turn results in SNc neurons degeneration due to excitotoxicity.

In order to examine the LDOPA role in the degeneration of SNc neurons in PD, we administer LDOPA after a certain percentage of SNc neuronal loss due to energy deficiency and investigate how LDOPA changes the course of SNc cell loss.

## 2.1. Izhikevich (Spiking) Neuronal Model (STN, GPe, MSN, CTX)

The Izhikevich neuronal models are capable of exhibiting biologically realistic firing patterns at a relatively low computational expense (Izhikevich, 2003). The proposed model of LIT consists of MSN (D1-MSN (G) and D1-MSN (GS)), STN, GPe, and CTX are modeled as spiking neuronal models using Izhikevich neuronal models arranged in the form of two-dimensional lattices (Figure 1). Based on the anatomical data of the rat basal ganglia (*Supplementary material-1*), the neuronal population sizes in the model were selected (Arbuthnott and Wickens, 2007; Oorschot, 1996). The Izhikevich parameters for MSN were adapted from (Humphries et al., 2009); for STN and GPe they were adapted from (Michmizos and Nikita, 2011) and (Mandali et al., 2015) respectively and those of CTX were adapted from (Izhikevich, 2003) are given in *Supplementary material-2*. The external bias current ($I^x$) was adjusted to match the firing rate of nuclei with published data (Tripathy et al., 2015).

The Izhikevich neuronal model of STN, GPe, CTX, and MSN consists of two variables, membrane potential ($v^x$) and membrane recovery variable ($u^x$):

$$C^x * \frac{d(v_{ij}^x)}{dt} = \varphi_1 - u_{ij}^x + I_{ij}^x + I_{ij}^{syn} \tag{1}$$

$$\frac{d(u_{ij}^x)}{dt} = a(b * \varphi_2 - u_{ij}^x) \tag{2}$$

$$if\ x = STN\ or\ GPe\ or\ CTX, \quad then$$

$$\varphi_1 = 0.04(v_{ij}^x)^2 + 5v_{ij}^x + 140 \tag{3}$$

$$\varphi_2 = v_{ij}^x$$

$$if\ x = MSN, \quad then$$

$$\varphi_1 = k^x(v_{ij}^x - v_r^x)(v_{ij}^x - v_t^x) \tag{4}$$

$$\varphi_2 = (v_{ij}^x - v_r^x)$$

Resetting:

$$if\ v_{ij}^x \geq v_{peak}^x, \quad then \quad \begin{Bmatrix} v_{ij}^x \leftarrow c \\ u_{ij}^x \leftarrow u_{ij}^x + d \end{Bmatrix} \tag{5}$$

where, $v_{ij}^x$, $u_{ij}^x$, $I_{ij}^x$, and $I_{ij}^{syn}$ are the membrane potential, the membrane recovery variable, the external bias current, and total synaptic current received to neuron $x$ at the location $(i,j)$ respectively; $v_t^x$ and $v_r^x$ are the threshold and resting potentials respectively; $k^x$ is the membrane constant, $C^x$ is the membrane capacitance, $\{a, b, c, d\}$ are Izhikevich parameters; $v_{peak}^x$ is the maximum membrane voltage set to neuron with $x$ being GPe or CTX or STN or MSN neuron.

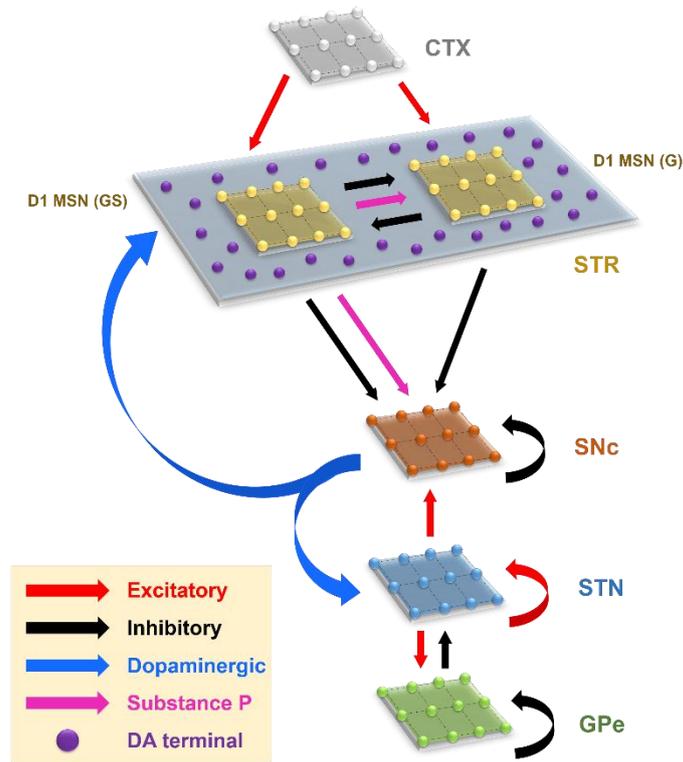

*Figure 1: Model architecture of the levodopa-induced toxicity. CTX, cortex; STR, striatum; D1 MSN (GS), D1-type medium spiny neuron (GABAergic and Substance P); D1 MSN (G), D1-type medium spiny neuron (GABAergic); SNc, substantia nigra pars compacta; STN, subthalamic nucleus; GPe, globus pallidus externa; DA, dopamine.*

**2.2.  Biophysical (Conductance-based) Neuronal Model (SNc soma)**

The biophysical neuronal model of SNc in the proposed LIT model was adapted from (Muddapu and Chakravarthy, 2020). The detailed biophysical model of SNc neuron consists of soma and terminal which includes cellular and molecular processes such as ion channels (including active ion pumps, ion exchangers), calcium buffering mechanism (calcium-binding proteins, calcium sequestration organelles (such as endoplasmic reticulum, mitochondria)), energy metabolism (glycolysis and oxidative phosphorylation), DA turnover processes (synthesis, storage, release, reuptake, and metabolism), molecular pathways involved in PD pathology (ROS formation and alpha-synuclein aggregation), and apoptotic pathways. The dynamics of SNc membrane potential ($v^{SNc}$) is given as,

$$\frac{d(v_{ij}^{SNc})}{dt} = \frac{F * v_{cyt}}{C^{SNc} * \mathcal{A}_{pmu}} * [J_{m,Na} + 2 * J_{m,Ca} + J_{m,K} + J_{inp}] \quad (6)$$

where, $F$ is the Faraday's constant, $C^{SNc}$ is the SNc membrane capacitance, $v_{cyt}$ is the cytosolic volume, $\mathcal{A}_{pmu}$ is the cytosolic area, $J_{m,Na}$ is the sodium membrane ion flux, $J_{m,Ca}$ is the calcium membrane ion flux, $J_{m,K}$ is the potassium membrane ion flux, $J_{inp}$ is the overall input current flux. The detailed information about the SNc neuron model was provided in (Muddapu and Chakravarthy, 2020).

## 2.3. Biochemical Dopamine Terminal Model (SNc terminal)

The biochemical DA terminal of SNc in the proposed LIT model was adapted from (Muddapu and Chakravarthy, 2020). The biochemical model of DA terminal consists of DA turnover processes, energy metabolism, and molecular pathways involved in PD pathology. The terminal is divided into two compartments, namely intracellular (cytoplasmic and vesicular) and extracellular compartments. The DA dynamics in the extracellular compartment ($[DA_e]$) was modeled as,

$$\frac{d([DA_e])}{dt} = J_{rel} - J_{DAT} - J_{eda}^o \quad (7)$$

where, $J_{rel}$ represents the flux of calcium-dependent DA release from the DA terminal, $J_{DAT}$ represents the unidirectional flux of DA translocated from the extracellular compartment (ECS) into the intracellular compartment (cytosol) via DA plasma membrane transporter (DAT), and $J_{eda}^o$ represents the outward flux of DA degradation which clears DA from ECS.

The DA dynamics in the intracellular compartment ($[DA_i]$) was modeled as

$$\frac{d([DA_i])}{dt} = \frac{d([DA_c])}{dt} + \frac{d([DA_v])}{dt} \quad (8)$$

where, $[DA_c]$ and $[DA_v]$ refer to the DA concentrations in the cytosolic and vesicular compartments, respectively.

The DA dynamics in the cytosolic compartment ($[DA_c]$) is given by,

$$\frac{d([DA_c])}{dt} = J_{DAT} - J_{VMAT} - J_{cda}^o + J_{ldopa} \quad (9)$$

where, $J_{DAT}$ represents the unidirectional flux of DA translocated from ECS into the cytosol through DAT, $J_{VMAT}$ represents the flux of DA into vesicle through vesicular monoamine transporters (VMAT), $J_{ida}^o$ represents the outward flux of DA degradation which clears DA from the cytosol, $J_{ldopa}$ represents the flux of synthesized cytosol DA from levodopa which is induced by calcium.

The DA dynamics in the vesicular compartment ($[DA_v]$) is given by,

$$\frac{d([DA_v])}{dt} = J_{VMAT} - J_{rel} \quad (10)$$

where, $J_{rel}$ represents the flux of calcium-dependent DA release from the DA terminal, $J_{VMAT}$ represents the flux of DA stored into a vesicle.

Based on the membrane activity, the DA turnover and other molecular processes were modulated in the terminal. The modulation of neuronal activity on the terminal was carried on by calcium dynamics, where calcium modulates DA synthesis and release. The calcium-induced synthesis of DA is given as,

$$J_{ldopa} = f([Ca_i]) \quad (11)$$

The calcium-induced release of DA is given as,

$$J_{rel} = f([Ca_i]) \quad (12)$$

where, $[Ca_i]$ is the intracellular calcium concentration in the DA terminal. For more details of the SNc terminal model, the reader may refer to (Muddapu and Chakravarthy, 2020).

## 2.4. Synaptic Connections

The synaptic connectivity among different neuronal populations was modelled as a standard single exponential model of postsynaptic currents (Humphries et al., 2009) as follows:

$$\tau_{Recep} * \frac{d(h_{ij}^{x \to y})}{dt} = -h_{ij}^{x \to y} + S_{ij}^{x}(t) \tag{13}$$

$$I_{ij}^{x \to y}(t) = W_{x \to y} * h_{ij}^{x \to y}(t) * (E_{Recep} - v_{ij}^{y}(t)) \tag{14}$$

The N-Methyl-D-aspartic Acid (NMDA) current was regulated by voltage-dependent magnesium channels which were modelled as,

$$B_{ij}(v_{ij}) = \frac{1}{1 + \left(\frac{[Mg^{2+}]}{3.57} * e^{-0.062 * v_{ij}^{y}(t)}\right)} \tag{15}$$

where, $h_{ij}^{x \to y}$ is the gating variable for the synaptic current from $x$ to $y$, $\tau_{Recep}$ is the decay constant for the synaptic receptor, $S_{ij}^{x}$ is the spiking activity of neuron $x$ at time $t$, $W_{x \to y}$ is the synaptic weight from neuron $x$ to $y$, $v_{ij}^{y}$ is the membrane potential of the neuron $y$ for the neuron at the location $(i, j)$, $E_{Recep}$ is the receptor-associated synaptic potential ($Recep$ = NMDA/AMPA/GABA), and $[Mg^{2+}]$ is the magnesium ion concentration. The time constants of Gamma-Amino Butyric Acid (GABA), Alpha-amino-3-hydroxy-5-Methyl-4-isoxazole Propionic Acid (AMPA) and NMDA in GPe, CTX, MSN, SNc, and STN were chosen from (Götz et al., 1997) as given in *Supplementary material-2*.

To accommodate extensive axonal arborization of SNc neurons (Bolam and Pissadaki, 2012), we considered one-to-many projections from SNc soma to SNc terminals (*Supplementary material-3*). The connectivity patterns among different neuronal populations were given in *Supplementary material-3*.

## 2.5. Lateral Connections

The lateral connections in SNc, STN and GPe, were modelled as Gaussian neighborhoods (Muddapu et al., 2019),

$$w_{ij,pq}^{m \to m} = A_m * e^{\frac{-d_{ij,pq}^{2}}{R_m^2}} \tag{16}$$

$$d_{ij,pq}^{2} = (i - p)^2 + (j - q)^2 \tag{17}$$

where, $w_{ij,pq}^{m \to m}$ is the lateral connection weight of neuron type $m$ at the location $(i,j)$, $d_{ij,pq}$ is the distance from center neuron $(p, q)$, $R_m$ is the variance of Gaussian, $A_m$ is the strength of lateral synapse, $m = GPe$ or $STN$ or $SNc$.

The connections within SNc and GPe populations were considered as inhibitory and within STN as excitatory (Muddapu et al., 2019) (Figure 1). No lateral connections were considered for both the MSNs and CTX populations. The lateral currents in the STN and GPe were modelled similar to equations (13-15) and in the case of SNc which was modelled as,

$$H_\infty = \frac{1}{1 + e^{\left(\frac{-\left(v_{ij}^x - \theta_g - \theta_g^H\right)}{\sigma_g^H}\right)}} \tag{18}$$

$$\frac{d(s_{ij}^{x \to y})}{dt} = \alpha * \left(1 - s_{ij}^{x \to y}\right) * H_\infty - \beta * s_{ij}^{x \to y} \tag{19}$$

$$I_{ij}^{x \to y}(t) = W_{x \to y} * s_{ij}^{x \to y} * \left(v_{ij}^y(t) - E_{GABA}\right) \tag{20}$$

where, $I_{ij}^{x \to y}$ is the synaptic current from neuron $x$ to $y$, $W_{x \to y}$ is the synaptic conductance from neuron $x$ to $y$, $v_{ij}^x$ and $v_{ij}^y$ are the membrane potential of the neuron $x$ and $y$ respectively for the neuron at the location $(i, j)$, $E_{GABA}$ is the GABAergic receptor potential, $s_{ij}^{x \to y}$ is the synaptic gating variable for the neuron. The parametric values of $\alpha$, $\beta$, $\theta_g$, $\theta_g^H$, $\sigma_g^H$ were adapted from (Rubin and Terman, 2004) and given in *Supplementary material-4*.

## 2.6. Neuromodulatory Effect on the Neuronal Populations

The effect of neuromodulators such as DA and substance P (SP) in the proposed LIT model was modelled based on (Muddapu et al., 2019) and (Buxton et al., 2017), respectively.

### 2.6.1. Dopaminergic modulation:

DA-modulated lateral connection strength in SNc, STN, and GPe populations. As DA level increases, the lateral connection strength in SNc and GPe increases, whereas, in the case of STN, it decreases. DA-modulation of lateral connection strength was modelled as,

$$A^{STN} = s_{max}^{STN} * e^{(-cd_{stn} * DA_S(t))} \tag{21}$$

$$A^{GPe} = s_{min}^{GPe} * e^{(cd_{gpe} * DA_s(t))} \qquad (22)$$

$$A^{SNc} = s_{min}^{SNc} * e^{(cd_{snc} * DA_s(t))} \qquad (23)$$

where, $s_{max}^{STN}$, $s_{min}^{GPe}$, and $s_{min}^{SNc}$ are lateral connection strengths at the basal spontaneous activity of the population without any external dopaminergic influence in $STN$, $GPe$, and $SNc$, respectively. $cd_{stn}$, $cd_{gpe}$, and $cd_{snc}$ were the factors by which DA affects the lateral connections in $STN$, $GPe$, and $SNc$ populations respectively, $DA_s(t)$ is the instantaneous DA level, which is the spatial average DA concentration of all the terminals. All parameter values are given in *Supplementary material-4*.

The post-synaptic effect of DA in SNc, STN and GPe was modelled similar to (Muddapu et al., 2019),

$$W_{x \to y} = (1 - cd2 * DA_s(t)) * w_{x \to y} \qquad (24)$$

where, $w_{x \to y}$ is the synaptic weight ($STN \to GPe, GPe \to STN, STN \to STN, GPe \to GPe, STN \to SNc, SNc \to SNc, MSN \to SNc$), $cd2$ is the parameter that affects the post-synaptic current, $DA_s(t)$ is the instantaneous DA level.

The effect of DA in the MSN population occurs on both synaptic and intrinsic ion channels (Surmeier et al., 2007). The cortical inputs to MSN were modulated by DA as similar to (Humphries et al., 2009),

$$I_{DA}^x(t) = I_{CTX \to MSN}^x(t) * \left(1 + \left(\frac{\beta_{DA}}{\alpha_{DA}^y}\right) * DA_s(t)\right) \qquad (25)$$

where, $I_{CTX \to MSN}^x$ is the synaptic current from $CTX$ to $MSN$ (where $x = NMDA$ or $AMPA$), $DA_s(t)$ is the instantaneous DA level, $\alpha_{DA}^y$ is the DA effect on $y$ neuron (where $y = D1 - MSN (GS)$ or $D1 - MSN (G)$), $\beta_{DA}$ was adapted from (Humphries et al., 2009).

In addition to modulating cortical afferent connections, DA also has effects on the intrinsic ion channels (Humphries et al., 2009) which was modelled in Izhikevich neuronal model as,

$$v_r^{DA} = v_r^{MSN} * \left(1 + K^{MSN} * \left(\frac{DA_s(t)}{\alpha_{DA}^y}\right)\right) \qquad (26)$$

$$d_{msn}^{DA} = d_{msn} * \left(1 - L^{MSN} * \left(\frac{DA_s(t)}{\alpha_{DA}^y}\right)\right) \qquad (27)$$

where, $v_r^{DA}$ and $d_{msn}^{DA}$ are the DA-modulated resting potential and after-spike reset value of $MSN$ respectively, $v_r^{MSN}$, and $d_{msn}$ are the resting potentials and after-spike reset value of $MSN$ respectively, $DA_s(t)$ is the instantaneous DA level; $\alpha_{DA}^y$ is the DA effect on $y$ neuron (where $y = D1 - MSN\ (GS)\ or\ D1 - MSN\ (G)$), $K^{MSN}$ and $L^{MSN}$ were adapted from (Humphries et al., 2009).

### 2.6.2. Substance P modulation:

SP modulates excitatory afferent connections of SNc (soma) and D1 MSN (G) in the proposed LIT model (Figure 1). It was observed that SP modulates the glutamatergic afferents of MSNs directly (Blomeley and Bracci, 2008) or indirectly (Blomeley et al., 2009) by co-release of SP by GABAergic D1 MSNs (Buxton et al., 2017; Reiner et al., 2010). In the proposed LIT model, we modelled SP-modulation of glutamatergic afferents of the D1 MSN (G) population by the D1 MSN (GS) population similar to (Buxton et al., 2017). It was observed that SP and tachykinin NK1 receptor (NK1-R) are highly expressed within the SNc (Lessard and Pickel, 2005; Mantyh et al., 1984; Ribeiro-da-Silva and Hökfelt, 2000; Sutoo et al., 1999; Thornton and Vink, 2015). SP-containing striatal neurons project to dopaminergic neurons where SP potentiates the release of striatal DA (Brimblecombe and Cragg, 2015; Thornton and Vink, 2015). It was reported that DA-dependent decrease in SP levels was observed in the basal ganglia regions (Sivam, 1991; Thornton et al., 2010; Thornton and Vink, 2015). Therefore, there exists a feedback regulation between DA and SP, which helps in maintaining DA homeostasis (Thornton et al., 2010; Thornton and Vink, 2015). In the proposed LIT model, we assumed that SP modulates STN glutamatergic inputs to SNc such that increased SP levels lead to excitation of SNc, which, in turn, enhances striatal DA level, modelled similar to (Buxton et al., 2017). Also, we incorporated SP-DA feedback regulation (SDFR) in SP-modulation in the proposed LIT model. The SP-modulation of glutamatergic inputs to D1 MSN (G) and SNc along with SDFR was given as,

$$I_{ij}^{x \to y}(t) = W_{x \to y} * h_{ij}^{x \to y}(t) * NSP * SDFR * \left(v_{ij}^y(t) - E_{Recep}\right) \qquad (28)$$

$$NSP = \left(1 + w_{sp} * N_{ij}^{sp}\left(t - \tau_d^{sp}\right)\right) \tag{29}$$

$$N_{ij}^{sp}(t) = \beta_{sp} * \left[1 - e^{\left(\frac{-A_{ij}^{sp}(t)}{\lambda_{sp}}\right)^{b_{sp}}}\right] \tag{30}$$

$$A_{ij}^{sp}(t) = \left[e^{\left(\frac{-S_{ij}^x(t)}{\tau_f^{sp}}\right)} - e^{\left(\frac{-S_{ij}^x(t)}{\tau_r^{sp}}\right)}\right] \tag{31}$$

$$SDFR = \left(1 - DA_s(t)\right) \tag{32}$$

where, $h_{ij}^{x \to y}$ is the gating variable for the synaptic current from $x$ to $y$, $W_{x \to y}$ is the synaptic weight from neuron $x$ to $y$, $S_{ij}^x$ is the spiking activity of neuron $x$ at time $t$, $v_{ij}^y$ is the membrane potential of the neuron $y$ for the neuron at the location $(i, j)$, $E_{Recep}$ is the receptor-associated synaptic potential ($Recep$ = NMDA/AMPA), $\tau_d^{sp}$ is the fixed time delay between MSN activity and the onset of neuropeptide effect, $\beta_{sp}$ is the gain factor, $N_{ij}^{sp}$ is the modulatory effect of SP, $w_{sp}$ is the influence of SP on $w_{STN \to SNc}$, $A_{ij}^{sp}$ is the amplitude of SP released which is induced by spiking activity $(S_{ij}^x)$, $DA_s(t)$ is the instantaneous DA level, $b_{sp}$ and $\lambda_{sp}$ were adapted from (Buxton et al., 2017) and given in *Supplementary material-4*.

## 2.7. Total Synaptic Current Received by Each Neuronal Type
### 2.7.1. SNc:

The total synaptic current received by a $SNc$ neuron at the lattice position $(i, j)$ is the summation of the glutamatergic input from the $STN$ neurons, considering both $NMDA$ and $AMPA$ receptor activation, comprising the GABAergic inputs from the $D1 - MSN\ (GS)$ and $D1 - MSN\ (G)$ neurons and lateral GABAergic current from other $SNc$ neurons.

$$\begin{aligned}I_{ij}^{SNcsyn} = &F_{STN \to SNc} * \left(I_{ij}^{NMDA \to SNc} + I_{ij}^{AMPA \to SNc}\right) \\ &+ \left(F_{D1-MSN\ (G) \to SNc} * I_{ij}^{D1-MSN(G) \to SNc}\right) \\ &+ \left(F_{D1-MSN\ (GS) \to SNc} * I_{ij}^{D1-MSN(GS) \to SNc}\right) + I_{ij}^{GABAlat}\end{aligned} \tag{33}$$

where, $I_{ij}^{NMDA \rightarrow SNc}$ and $I_{ij}^{AMPA \rightarrow SNc}$ are the glutamatergic currents corresponding to $NMDA$ and $AMPA$ receptors activation respectively; $I_{ij}^{D1-MSN(G) \rightarrow SNc}$ and $I_{ij}^{D1-MSN(GS) \rightarrow SNc}$ are the GABAergic inputs from the $D1-MSN\ (G)$ and $D1-MSN\ (GS)$ neurons respectively; $I_{ij}^{GABAlat}$ is the lateral GABAergic current from other $SNc$ neurons; $F_{STN \rightarrow SNc}$ is the scaling factor for the glutamatergic current from $STN$ neuron; $F_{D1-MSN\ (G) \rightarrow SNc}$ is the scaling factor for the GABAergic current from $D1-MSN\ (G)$ neuron; $F_{D1-MSN\ (GS) \rightarrow SNc}$ is the scaling factor for the GABAergic current from $D1-MSN\ (GS)$ neuron.

### 2.7.2. GPe:

The total synaptic current received by a $GPe$ neuron at the lattice position $(i, j)$ is the summation of the glutamatergic input from the $STN$ neurons considering both $NMDA$ and $AMPA$ receptors activation and the lateral GABAergic current from other $GPe$ neurons.

$$I_{ij}^{GPesyn} = I_{ij}^{NMDA \rightarrow GPe} + I_{ij}^{AMPA \rightarrow GPe} + I_{ij}^{GABAlat} \quad (34)$$

where, $I_{ij}^{NMDA \rightarrow GPe}$ and $I_{ij}^{AMPA \rightarrow GPe}$ are the glutamatergic currents from $STN$ neuron considering both $NMDA$ and $AMPA$ receptors activation respectively; $I_{ij}^{GABAlat}$ is the lateral GABAergic current from other $GPe$ neurons.

### 2.7.3. STN:

The total synaptic current received by a $STN$ neuron at the lattice position $(i, j)$ is the summation of the GABAergic input from the $GPe$ neurons and the lateral glutamatergic input from other $STN$ neurons considering both $NMDA$ and $AMPA$ receptors activation.

$$I_{ij}^{STNsyn} = I_{ij}^{GABA \rightarrow STN} + I_{ij}^{NMDAlat} + I_{ij}^{AMPAlat} \quad (35)$$

where, $I_{ij}^{GABA \rightarrow STN}$ is the GABAergic current from $GPe$ neuron; $I_{ij}^{NMDAlat}$ and $I_{ij}^{AMPAlat}$ are the lateral glutamatergic current from other $STN$ neurons considering both $NMDA$ and $AMPA$ receptors activation, respectively.

### 2.7.4. D1-MSN (GS):

The total synaptic current received by a $D1-MSN\,(GS)$ neuron at the lattice position $(i,j)$ is the summation of the GABAergic input from the $D1-MSN\,(G)$ neurons and the glutamatergic input from $CTX$ neurons considering both $NMDA$ and $AMPA$ receptors activation.

$$I_{ij}^{D1-MSN(GS)syn} = I_{ij}^{GABA \to D1-MSN(GS)} + I_{ij}^{NMDA \to D1-MSN(GS)} + I_{ij}^{AMPA \to D1-MSN(GS)} \tag{36}$$

where, $I_{ij}^{GABA \to D1-MSN(GS)}$ is the GABAergic current from $D1-MSN(G)$ neuron, $I_{ij}^{NMDA \to D1-MSN(GS)}$, and $I_{ij}^{AMPA \to D1-MSN(GS)}$ are the glutamatergic current from $CTX$ neurons considering both $NMDA$ and $AMPA$ receptors activation, respectively.

### 2.7.5.   D1-MSN (G):

The total synaptic current received by a $D1-MSN\,(G)$ neuron at the lattice position $(i,j)$ is the summation of the GABAergic input from the $D1-MSN\,(GS)$ neurons and the glutamatergic input from $CTX$ neurons considering both $NMDA$ and $AMPA$ receptor activation.

$$I_{ij}^{D1-MSN(G)syn} = I_{ij}^{GABA \to D1-MSN(G)} + I_{ij}^{NMDA \to D1-MSN(G)} + I_{ij}^{AMPA \to D1-MSN(G)} \tag{37}$$

where, $I_{ij}^{GABA \to D1-MSN(G)}$ is the GABAergic current from $D1-MSN(GS)$ neuron, $I_{ij}^{NMDA \to D1-MSN(G)}$, and $I_{ij}^{AMPA \to D1-MSN(G)}$ are the glutamatergic current from $CTX$ neurons considering both $NMDA$ and $AMPA$ receptors activation, respectively.

### 2.8.   Neurodegeneration of SNc neurons

Calcium plays a dual role in living organisms as a survival factor or a ruthless killer (Orrenius et al., 2003). For the survival of neurons, minimal (physiological) levels of glutamate stimulation are required. Under normal conditions, calcium concentration within a cell is tightly regulated by pumps, transporters, calcium-binding proteins, endoplasmic reticulum (ER) and mitochondria (Surmeier et al., 2011; Wojda et al., 2008). Due to prolonged calcium influx driven by excitotoxicity, the calcium homeostasis within the cell is disturbed, which results in cellular imbalance, leading to activation of apoptotic pathways (Bano and Ankarcrona, 2018). The SNc soma undergoes degeneration when calcium builds up inside the cell becomes high, resulting in calcium loading inside ER and mitochondria, which leads to ER-stress-induced and mitochondrial-induced apoptosis respectively (Malhotra and Kaufman, 2011). In the proposed LIT model, we incorporate a mechanism of programmed cell death, whereby a SNc neuron under high stress (high calcium levels) kills itself. The stress in a given SNc neuron was observed by monitoring the intracellular calcium concentrations in the cytoplasm, ER, and mitochondria.

The SNc neuron undergoes ER-stress-induced apoptosis when calcium levels in ER cross a certain threshold ($ER_{thres}$). Under such conditions, the particular SNc neuron gets eliminated as follows,

$$if \quad Ca_{ij}^{ER}(t) > ER_{thres}, \quad then \quad v_{ij}^{SNc}(t) = 0 \qquad (38)$$

where, $Ca_{ij}^{ER}$ is the calcium concentration in the ER, $ER_{thres}$ is the calcium concentration threshold after which ER-stress induced apoptosis gets initiated, $v_{ij}^{SNc}$ is the membrane voltage of neuron at the lattice position $(i, j)$.

The SNc neuron undergoes mitochondria-induced apoptosis when calcium levels in mitochondria cross a certain threshold ($MT_{thres}$). Then that particular SNc neuron will be eliminated as follows,

$$if \quad Ca_{ij}^{MT}(t) > MT_{thres}, \quad then \quad v_{ij}^{SNc}(t) = 0 \qquad (39)$$

where, $Ca_{ij}^{MT}$ is the calcium concentration in mitochondria, $MT_{thres}$ is the calcium concentration threshold after which mitochondria-induced apoptosis gets initiated, $v_{ij}^{SNc}$ is the membrane voltage of neuron at the lattice position $(i, j)$.

When calcium concentration in ER crosses a certain threshold, there is an efflux of excess calcium from ER out into cytoplasm, which in turn activates calpain, resulting in activation of proapoptotic factors through cytochrome-c independent apoptotic pathway. Similarly, when calcium concentration in MT crosses a certain threshold, excess calcium in MT results in the formation of mitochondrial transition pores (MTPs). Proapoptotic cytochrome-c released from MT through MTPs, which triggers cytochrome-c dependent apoptosis. In the proposed modeling study, when the apoptotic signal gets activated from either of the pathways in a particular neuron, we formulate an approach wherein that particular neuron was eliminated by making $v_{ij}^{SNc}(t) = 0$ from the time $t$ till the end of the simulation.

## 2.9. Terminal degeneration of SNc neurons

DA is the primary contributor to the oxidative stress in the neuron (Lotharius et al., 2005; Luo and Roth, 2000; Miyazaki and Asanuma, 2008). To evade oxidative stress, SNc neurons tightly regulate the DA turnover processes (Guo et al., 2018). It was suggested that methamphetamine-induced dopaminergic nerve terminal loss (Ares-Santos et al., 2014; Cadet et al., 2003; Ricaurte et al., 1982, 1984) is precipitated by oxidative stress (De Vito and Wagner, 1989) by enhancing cytoplasmic DA levels (Larsen et al., 2002; Mark et al., 2004). In the proposed LIT model, the oxidative stress in the SNc terminals was observed by monitoring intracellular reactive oxygen species (ROS) concentration. The SNc terminal is eliminated when ROS levels in the terminal cross certain threshold ($ROS_{thres}$) as follows,

$$if \quad ROS_{ij}^T(t) > ROS_{thres}, \quad then \quad Ca_{ij}^T(t) = 0 \qquad (40)$$

where, $ROS_{ij}^T$ is the ROS concentration in the SNc terminal, $ROS_{thres}$ is the ROS concentration threshold above which oxidative stress-induced terminal degeneration gets initiated; $Ca_{ij}^T$ is the calcium concentration of the SNc terminal at the lattice position $(i,j)$.

When ROS level crosses a certain threshold, excess ROS triggers degeneration of the terminal. In the proposed modeling study, when ROS level crosses the threshold in a particular terminal, we formulate an approach wherein that particular terminal was eliminated by making $Ca_{ij}^T(t) = 0$ from the time $t$ till the end of the simulation since calcium plays an important role in terminal functioning.

## 2.10. Neuroprotective Strategies
### 2.10.1. Levodopa Therapy

To alleviate PD symptoms, the most potent drug, LDOPA, a precursor of DA, is typically administrated (Jankovic and Aguilar, 2008). During medication, serum LDOPA is taken up from the blood into the extracellular fluid by aromatic L-amino acid transporter by competing with other amino acids (Camargo et al., 2014; Figura et al., 2018). L-DOPA, thus absorbed into the bloodstream, later enters SNc terminals and gets converted to DA by aromatic L-amino acid decarboxylase (Khor and Hsu, 2007). In the proposed LIT model, serum LDOPA uptake into SNc terminal from the blood was modelled as a single step along with competition with other amino acids such as tyrosine and tryptophan (Porenta and Riederer, 1982). It was represented using the Michaelis-Menten equation (Chou, 1976) where serum LDOPA competes with serum tyrosine and serum tryptophan for transporter (Reed et al., 2012) as given below:

$$V_{trans} = \frac{V_{trans}^{max} * [LDOPA_S]}{K_m^{LDOPA_S} * \left(1 + \frac{[TYR_s]}{K_a^{TYR_s}} + \frac{[TRP_s]}{K_a^{TRP_s}}\right) + [LDOPA_S]} \quad (41)$$

where, $V_{trans}^{max}$ is the maximum flux through aromatic L-amino acid transporter, $[LDOPA_S]$ is the serum LDOPA concentration, $K_m^{LDOPA_S}$ is the concentration of $[LDOPA_S]$ at which velocity of the transporter attained half of the maximal velocity, $[TYR_s]$ is the serum tyrosine concentration, $[TRP_s]$ is the serum tryptophan concentration. $K_a^{TYR_s}$ is the affinity constant for $[TYR_s]$, $K_a^{TRP_s}$ is the affinity constant for $[TRP_s]$.

LDOPA therapy was implemented in the proposed LIT model by the following criterion,

$$[LDOPA_s](N_{sc}^z, t) = \begin{cases} 0, & N_{sc}^z(t) > T_l^z \\ [LDOPA_s^{med}], & N_{sc}^z(t) \leq T_l^z \end{cases} \quad (42)$$

$$T_l^z = P_z^{snc} - (pcl * P_z^{snc}) \quad (43)$$

where, $[LDOPA_s](N_{sc}^z, t)$ is the instantaneous serum $[LDOPA]$ concentration based on the number of surviving SNc neurons or terminals at the time $(t)$ $(N_{sc}^z(t))$, $[LDOPA_s^{med}]$ is the serum $[LDOPA]$ concentration during medication, $N_{sc}^z(t)$ is the instantaneous number of surviving SNc neurons or terminals, $pcl$ is the percentage of SNc cell or terminal loss (25 %) at which therapeutic intervention was employed ($pcl = 0.25$), $T_l^z$ represents the number of surviving SNc cells or terminals at which therapeutic intervention was employed, $P_z^{snc}$ is the population size of $z$ ($z = soma\ or\ terminal$). In the present study, the therapeutic intervention is given at 25% SNc cell or terminal loss.

### 2.10.2. SP antagonist therapy

It was reported that SP exacerbated dopaminergic neurodegeneration in mice (Wang et al., 2014), and therefore administrating SP antagonists creates neuroprotection of dopaminergic neurons in PD (Johnson et al., 2017; Thornton and Vink, 2012, 2015). In the proposed LIT model, SP antagonist effect was implemented as,

$$w_{spa} = w_{sp} * \delta_{spa} \qquad (44)$$

where, $w_{sp}$ is the influence of SP on $w_{STN \rightarrow SNc}$, $\delta_{spa}$ is the proportion of SP inhibition, $w_{spa}$ is the influence of SP on $w_{STN \rightarrow SNc}$ under SP antagonist therapy.

The SP antagonist therapy was implemented in the proposed LIT model by the following criterion,

$$\delta_{spa}(N_{sc}^z, t) = \begin{cases} 0, & N_{sc}^z(t) > T_l^z \\ \delta_{spa}^{med}, & N_{sc}^z(t) \leq T_l^z \end{cases} \qquad (45)$$

where, $\delta_{spa}(N_{sc}^z, t)$ is the instantaneous proportion of SP inhibition based on the number of surviving SNc neurons or terminals at the time $(t)$ $(N_{sc}^z(t))$, $\delta_{spa}^{med}$ is the proportion of SP inhibition during therapy, $N_{sc}^z(t)$ is the instantaneous number of surviving SNc neurons or terminals, $T_l^z$ represents the number of surviving SNc cells or terminals at which therapeutic intervention was employed ($z = soma\ or\ terminal$).

### 2.10.3. Glutathione therapy

The impaired DA metabolism causes oxidative stress (ROS), which in turn leads to PD pathogenesis (Masato et al., 2019). It was reported that abnormal activity of vesicular monoamine transporter 2 (VMAT2) leads to reduced vesicular DA storage and increased cytoplasmic DA which results in oxidative stress-induced degeneration of cell bodies (soma) and terminals (Caudle et al., 2007; Kariya et al., 2005; Mingazov and Ugryumov, 2019; Pifl et al., 2014). It was reported that the glutathione (GSH) administration results in the improvement of PD symptoms, but the underlying mechanism is not clear (Hauser et al., 2009; Mischley et al., 2017; Zeevalk et al., 2008). We suggest that glutathione administration might result in scavenging of ROS, which in turn leads to neuroprotection (Li et al., 2015). In the proposed LIT model, glutathione effect was implemented as,

$$[GSH_{gst}^z] = [GSH^z] + [GSH_{gs}^z] \qquad (46)$$

where, $[GSH_{gs}^z]$ is the GSH concentration under glutathione therapy ($z = soma\ or\ terminal$), $[GSH^z]$ is the GSH concentration.

The glutathione therapy was implemented in the proposed LIT model by the following criterion,

$$[GSH_{gs}^z](N_{sc}^z, t) = \begin{cases} 0, & N_{sc}^z(t) > T_l^z \\ [GSH_{med}^z], & N_{sc}^z(t) \leq T_l^z \end{cases} \qquad (47)$$

where, $[GSH_{gs}^z](N_{sc}^z, t)$ is the instantaneous $[GSH]$ therapy based on the number of surviving SNc neurons or terminals at the time $(t)$ $(N_{sc}^z(t))$, $N_{sc}^z(t)$ is the instantaneous number of surviving SNc neurons or terminals, $[GSH_{med}^z]$ is the $[GSH]$ concentration dosage under $GSH$ therapy, $T_l^z$ represents the number of surviving SNc cells or terminals at which therapeutic intervention was employed ($z = soma\ or\ terminal$).

## 3. RESULTS

We investigate the Izhikevich models of the neurons of CTX, MSN, GPe, and STN, which were chosen from the literature (Humphries et al., 2009; Mandali et al., 2015; Michmizos and Nikita, 2011) for their characteristic firing pattern and other biological properties (Figure 2). Along with the above Izhikevich neuronal models, we also investigate the biophysical neuronal model of SNc for its characteristic responses (Figure 3). Next, we explored the effect of DA and SP on the network of MSN and SNc neurons and compared them with published data (Figure 4).

Then, we show the effect of homogeneous (Figure 5) and heterogeneous (Figure 6) energy deficit conditions on the progression of SNc somas and terminals loss. Next, we show the effect of extracellular LDOPA on the progression of SNc soma and terminal loss under energy deficit conditions (Figure 7A). Finally, we explore various therapeutics such as LDOPA (Figure 7B), SP antagonist (Figure 7C), and glutathione (Figure 7D) for their neuroprotective effect on the progression of SNc somas and terminals loss under energy deficit conditions.

## 3.1. Characteristic Firing Response of Different Neuronal Types

The response of a single neuronal model of five different neuronal types involved in the proposed LIT model for different external applied currents is shown in Figure 2. The basal firing frequencies of the different neuronal types were matched with experimentally observed firing (Tripathy et al., 2014) by adjusting $I_{ij}^x$ parameter given in *Supplementary material-2*.

The pyramidal neurons in the cortex are broadly classified into three types, namely regular spiking (RS), intrinsic (inactivating) bursting, and non-inactivating bursting neurons. The regular spiking neurons further subdivided into fast-adapting and slow-adapting types of neurons (Degenetais, 2002). The time-averaged firing rate of all neuronal types varies widely, ranging from $< 1\ Hz$ up to several tens of hertz (Griffith and Horn, 1966; Koch and Fuster, 1989). The spontaneous firing rates of all pyramidal cortical neuronal types are as follows: fast-adapting RS ($0.62 \pm 0.75\ Hz$), slow-adapting RS ($0.90 \pm 1.23\ Hz$), intrinsic bursting ($3.1 \pm 2.6\ Hz$) and non-inactivating bursting ($2.8 \pm 3.2\ Hz$) (Degenetais, 2002). In the proposed LIT model, we adjust $I_{ij}^x$ value such that the combined spontaneous and stimulus-driven firing rate of the pyramidal cortical neuron falls in the range of $10 - 15\ Hz$ (Figure 2A).

The medium spiny neurons in the striatum broadly classified into two types based on the type of DA receptor present, namely D1 and D2-types. In the proposed LIT model, we consider only D1-type MSNs because they only project GABAergic inputs to SNc neurons (Gerfen, 1985). The spontaneous firing rate of MSN was observed experimentally to be in the range of $0.6 - 16.1\ Hz$ (Mahon et al., 2006; Pitcher et al., 2007). In the proposed LIT model, we adjust $I_{ij}^x$ value such that the combined spontaneous and cortical-driven firing rate of MSN falls in the range of $10 - 20\ Hz$ (Figure 2B).

The STN neurons exhibit two distinct types of firing patterns experimentally: tonic pacemaking firing and phasic high-frequency bursting (Allers et al., 2003; Beurrier et al., 1999). The spontaneous firing rate of STN neurons was observed experimentally to be in the range of $6 - 30\ Hz$ (Allers et al., 2003; Lindahl et al., 2013). In the proposed LIT model, we adjust $I_{ij}^x$ value such that the STN spontaneous firing rate is $\sim 13\ Hz$ (Figure 2D).

The GPe neurons exhibit an atypical type of firing pattern where bursts and pauses appear aperiodically in a continuous tonic high-frequency firing (Hegeman et al., 2016; Kita and Kita, 2011). The spontaneous firing rate of GPe neuron was observed experimentally to be in the range of $8 - 60\ Hz$ (Bugaysen et al., 2010; Elias et al., 2008; Lindahl et al., 2013). In the proposed LIT model, we adjusted $I_{ij}^x$ value such that the GPe spontaneous firing rate was $\sim 30\ Hz$ (Figure 2C).

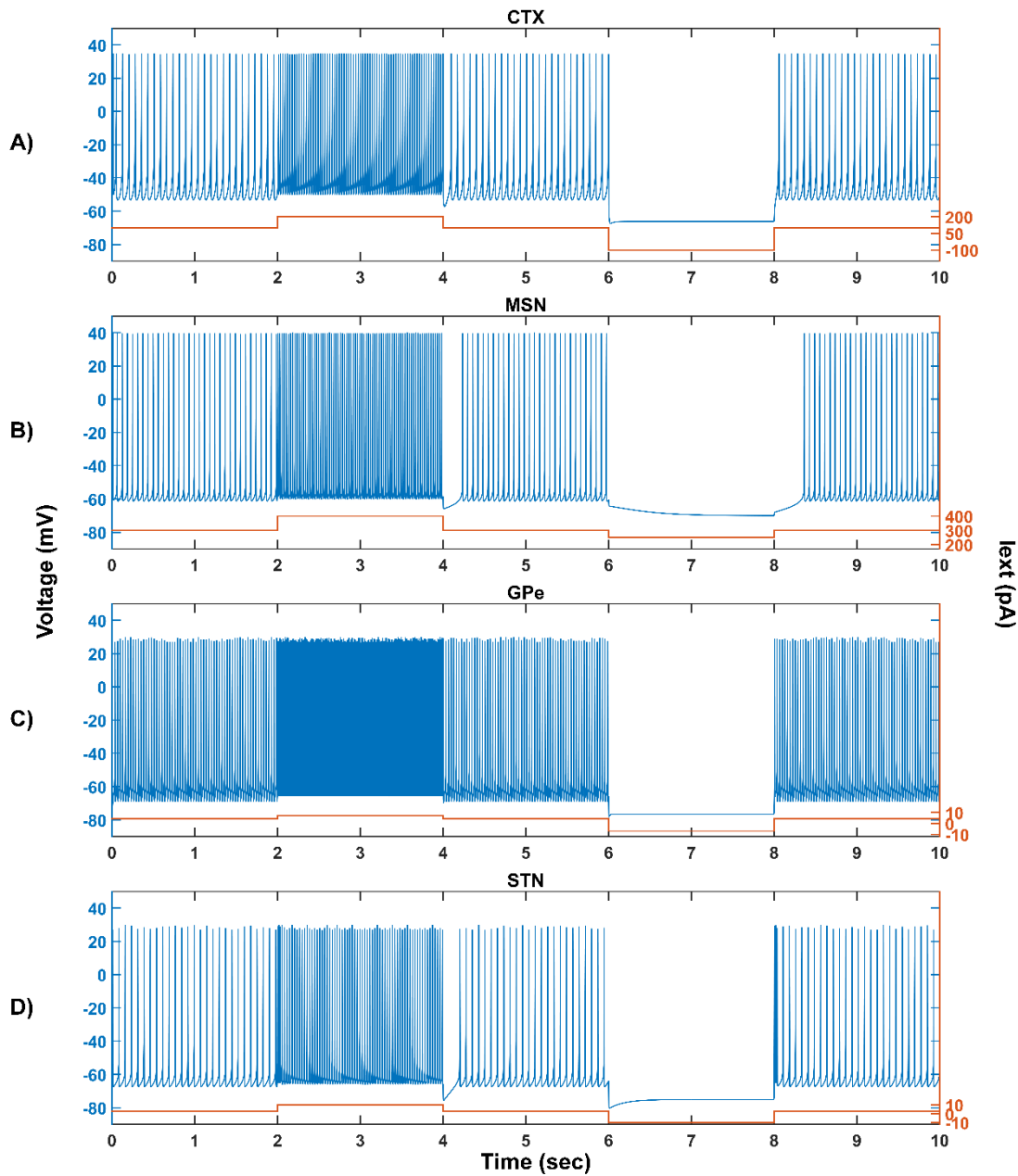

*Figure 2: Characteristic behavior of individual neurons. Characteristic behavior of individual CTX **(A)**, MSN **(B)**, GPe **(C)**, and STN **(D)** neuronal types. CTX, cortex; MSN, medium spiny neuron; GPe, globus pallidus externa; STN, subthalamic nucleus, Iext, external current; pA, picoampere; mV, millivolt; sec, second.*

The SNc neurons exhibit two distinct types of firing patterns experimentally, namely background or low-frequency irregular tonic firing $(3-8\ Hz)$ and bursting or high-frequency regular phasic firing $(\sim 20\ Hz)$ (Grace and Bunney, 1984b, 1984a). In the proposed LIT model, SNc neurons spontaneously fire with a firing rate of $\sim 4\ Hz$ (Figure 3). The calcium concentration inside the SNc neuron during the resting state was $\sim 1x10^{-4}\ mM$, and it rises up to $1x10^{-3}\ mM$ upon arrival of the action potential (Figure 3C) (Ben-Jonathan and Hnasko, 2001; Dedman and Kaetzel, 1997; Wojda et al., 2008). The dopamine released by SNc neuron during tonic spiking was $\sim 45\ x\ 10^{-6}\ mM$, which was in the range of $(34 - 48)\ x\ 10^{-6}\ mM$ observed experimentally (Figure 3D) (Garris et al., 1997). Upon injection of depolarizing external current (continuous pulse ($I_{ext} = 25\ x\ 10^{-6}\ pA$) and duration (1 sec)), SNc neuron exhibits bursting type of firing which lasts for more than one second after the pulse was removed (Figure 2B, positive current), demonstrating the slow-adapting nature of SNc neuron due to excess calcium build-up inside the neuron (Figure 3C, positive current) (Kuznetsova et al., 2010). During the depolarizing current stimulation, SNc neurons exhibit the property within a burst that spikes after an initial spike showed a decrease in amplitude (Figure 3B, positive current), which is a characteristic bursting property of SNc neurons (Grace and Bunney, 1984a). The dopamine concentration released by SNc neuron during phasic bursting peaked at $\sim 118\ x\ 10^{-6}\ mM$ (Figure 3D, positive current), which is in the range of $(90 - 220)\ x\ 10^{-6}\ mM$ (Chen and Budygin, 2007). Further increase in depolarizing current amplitude increases extracellular DA release exponentially but never exceeds beyond $1\ x\ 10^{-3}\ mM$ (not shown) (Gonon, 1988). Upon injection of hyperpolarized external current (continuous pulse ($I_{ext} = -300\ x\ 10^{-6}\ pA$) and duration (1 sec)), SNc neuron exhibits quiescent state until stimulation was removed (Figure 3B, negative current). Due to hyperpolarized current stimulation, the calcium oscillation in SNc neuron was minimal (Figure 3C, negative current), which was reflected in the near absence of extracellular DA release (Figure 3D, negative current).

The lateral connections in SNc, STN, and GPe neuronal populations were studied in the previous work (Muddapu et al., 2019). No lateral connections were considered in CTX and MSN neuronal populations for the simplification of the proposed LIT model.

## 3.2. Neuromodulatory Effect of Dopamine on MSN and SNc Neuronal Populations

In the MSN population, DA affects both synaptic and intrinsic ion channels (Surmeier et al., 2007). As the DA levels increase, the influence of cortical glutamatergic inputs on D1-type MSN increases resulting in monotonously increasing firing frequency (Figure 4A), which was consistent with experimental work (Cepeda et al., 1993) and previous modeling studies (Humphries et al., 2009). In the MSN population, SP affects synaptic ion channels, especially glutamatergic afferents (Blomeley et al., 2009; Blomeley and Bracci, 2008). As the SP levels (or SP scaling factor ($W_{sp}$)) increases, the influence of cortical glutamatergic inputs on D1-type MSN increases resulting in monotonously increasing firing frequency (Figure 4B), which was similar to experimental (Blomeley and Bracci, 2008) and other modelling studies (Buxton et al., 2017).

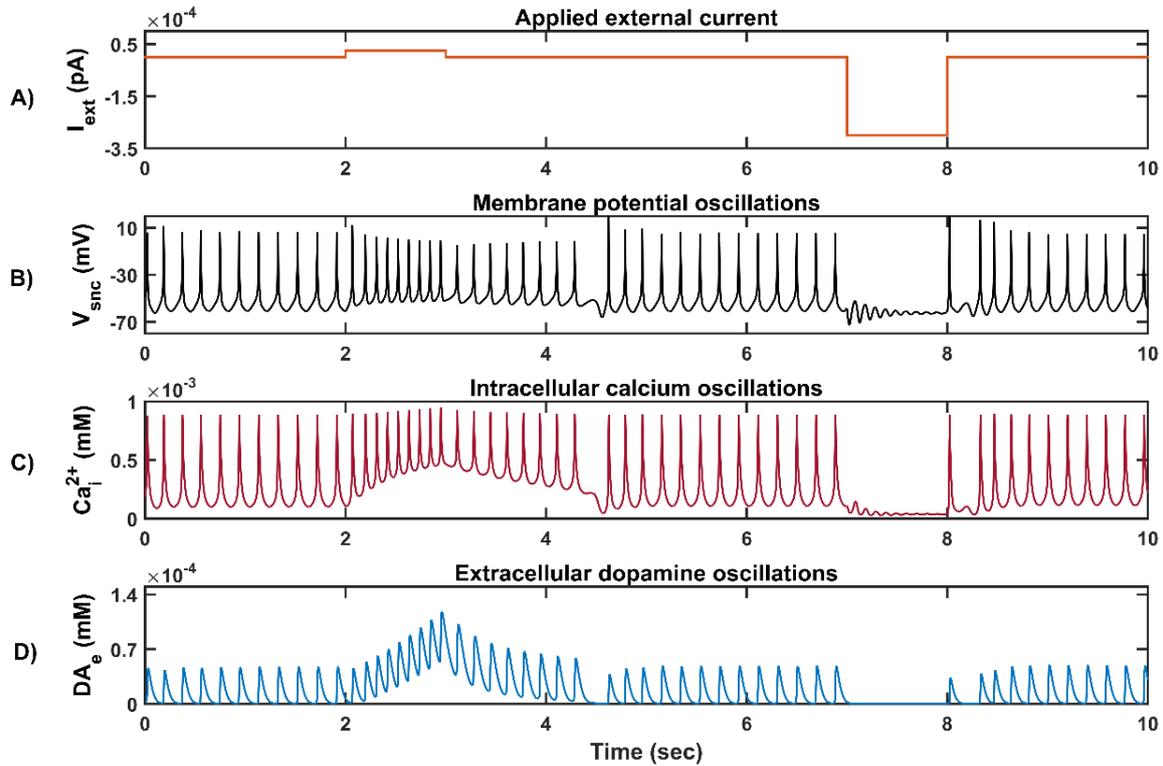

*Figure 3: Characteristic behavior of individual SNc neuron. (A) Applied external current ($I_{ext}$), (B) Membrane potential oscillations ($V_{snc}$), (C) Intracellular calcium oscillations ($Ca_i^{2+}$), (D) Extracellular dopamine concentration ($DA_e$). pA, picoampere; mV, millivolt; sec, second; mM, millimolar.*

In the SNc population, DA affects both synaptic and lateral connections (Muddapu et al., 2019). As the DA level increases, the influence of synaptic and lateral connection inputs on SNc increases resulting in non-monotonously decreasing firing frequency (Figure 4C) which was similar to experimental (Ford, 2014; Hebb and Robertson, 1999; Tepper and Lee, 2007; Vandecasteele et al., 2005) and other modelling studies (Muddapu et al., 2019). In the SNc population, SP affects the synaptic ion channels, especially glutamatergic afferents (Brimblecombe and Cragg, 2015; Thornton and Vink, 2015). As the SP level (or SP scaling factor ($W_{sp}$)) increases, the influence of STN glutamatergic inputs on SNc increases, resulting in monotonously increasing firing frequency (Figure 4D), which was similar to experimental studies (Brimblecombe and Cragg, 2015). The effect of DA on STN and GPe neuronal populations was simulated in the previous work (Muddapu et al., 2019).

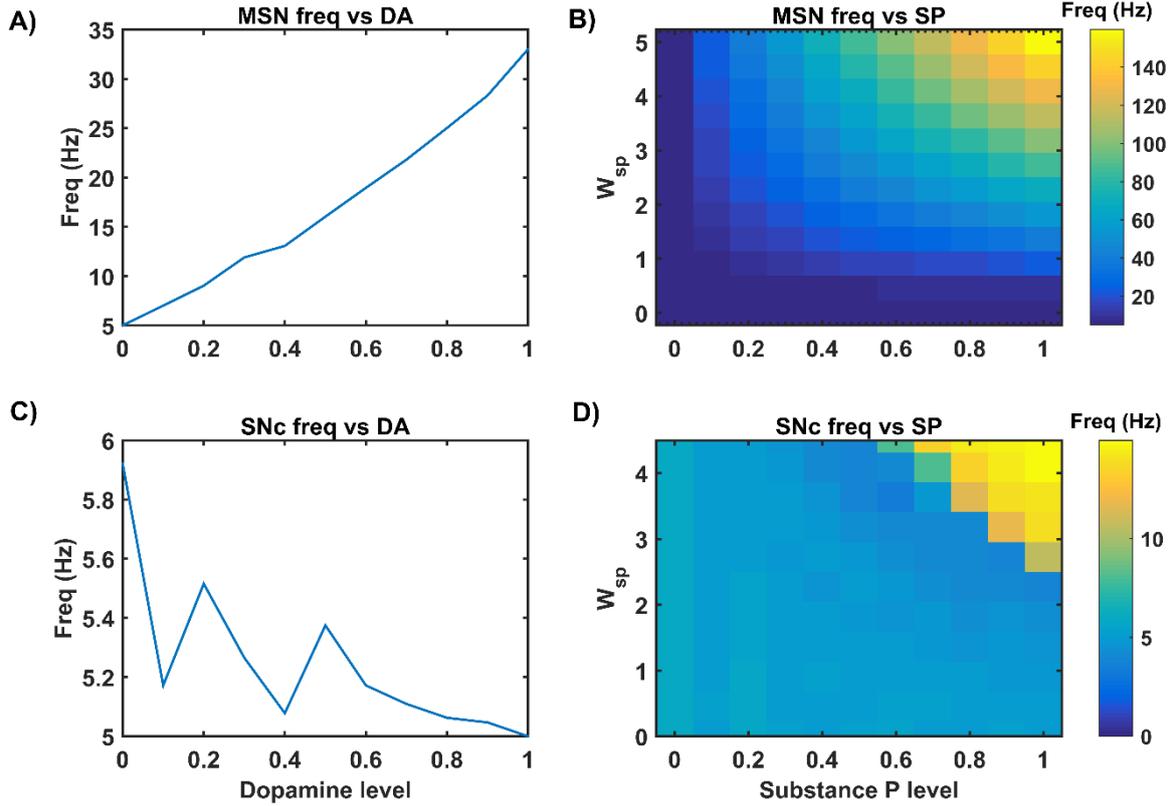

*Figure 4: Neuromodulatory effects of DA and SP on MSN and SNc neuronal populations. Effect of dopamine on MSN **(A)** and SNc **(C)** populations. Effect of SP on MSN **(B)** and SNc **(D)**. MSN, medium spiny neuron; DA, dopamine; Freq, frequency; SP, substance P; SNc, substantia nigra pars compacta; $W_{sp}$, scaling factor of SP influence; Hz, hertz.*

### 3.3. Energy Deficiency Occurring Similarly in SNc Somas and Terminals

In order to study energy deficiency as the possible root cause of SNc cell loss PD, we simulated ischemic conditions by modulating glucose and oxygen inputs in the proposed LIT model. As SNc somas and terminals are far from each other, the ischemic condition was implemented in two ways: homogeneous (energy deficiency occurs similarly in somas (midbrain) and terminals (striatum)) and heterogeneous (energy deficiency occurs differently in somas (midbrain) and terminals (striatum)). Homogeneous energy deficiency was implemented by reducing glucose and oxygen values by same proportions in both SNc somas and terminals. In the case of homogeneous energy deficiency, terminal loss starts with just 10% of somas and terminals in energy deficiency (Figure 5A, orange bar). However, soma loss starts at 70% of somas and terminals in energy deficiency (Figure 5A, blue bar). To observe the influence of STN on SNc soma and terminal loss, the currents from STN to SNc were monitored, which showed higher positive currents after 60% of somas and terminals are in energy deficiency (Figure 5B). The higher positive current from STN also results in increased ROS production in both SNc somas and terminals (Figure 5C).

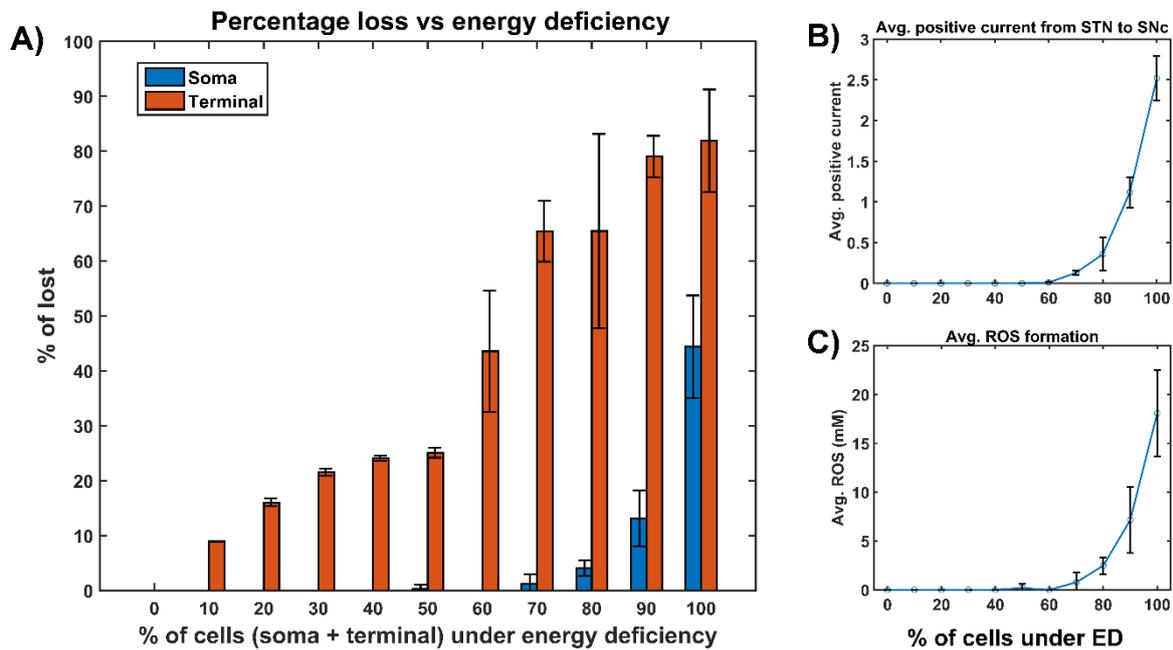

*Figure 5: The model response under homogeneous energy deficiency. (A) The percentage loss of SNc somas and terminals, (B) Average positive current from STN to SNc, (C) Average ROS formation. STN, subthalamic nucleus; SNc, substantia nigra pars compacta; ROS, reactive oxygen species; mM, millimolar; ED, energy deficiency.*

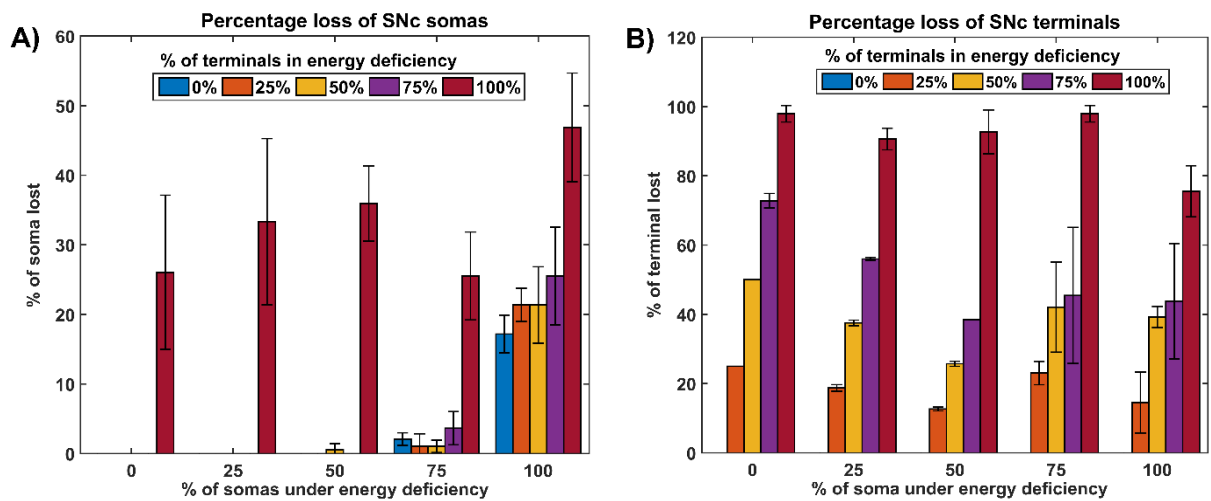

*Figure 6: The model response under heterogeneous energy deficiency. (A) The percentage loss of SNc somas, (B) The percentage loss of SNc terminals. ED, energy deficiency; SNc, substantia nigra pars compacta.*

### 3.4. Energy Deficiency Occurring Differently in SNc Somas and Terminals

Heterogeneous energy deficiency was implemented by reducing glucose and oxygen values by different proportions in both SNc somas and terminals. In the case of heterogeneous energy deficiency, the loss of somas was observed after 75% of somas are in energy deficiency for all energy deficiency levels in terminals (Figure 6A). The soma loss was minimal or non-existent when the percentage of somas in energy deficiency was below 75% for all levels of energy deficiency in terminals except 100%. In the case when 100% of terminals are in energy deficiency, the loss of somas was above 20% for all levels of energy deficiency in somas, and maximum loss of somas ($\sim 45\%$) was observed when 100% of somas are energy deficiency (Figure 6A). The terminal loss observed for all percentages of somas and terminals in energy deficiency except 0% of terminals in energy deficiency (Figure 6B). The terminal loss increases with an increase in the percentage of terminals in energy deficiency for all percentages of somas in energy deficiency (Figure 6B).

### 3.5. Effect of Extracellular LDOPA

In order to study the effect of extracellular (serum) LDOPA on SNc somas and terminal loss in energy deficiency (100 % energy deficiency), we have modified extracellular LDOPA concentration in the range from $36 \times 10^{-9}\ mM$ to $36\ mM$ with the multiple of 10x (Figure 7A). With lower concentrations of extracellular LDOPA ranging from $36 \times 10^{-9}\ mM$ to $36 \times 10^{-7}\ mM$, a more significant loss ($\sim 60\%$) of SNc somas was observed when compared to SNc terminals ($\sim 40\%$) (Figure 7A). With higher concentrations of extracellular LDOPA ranging from $36 \times 10^{-4}\ mM$ to $36\ mM$, more of SNc terminal loss ($\sim 95\%$) was observed when compared to SNc somas ($\sim 35\%$) (Figure 7B). At intermediate levels of extracellular LDOPA concentrations ranging from $36 \times 10^{-7}\ mM$ to $36 \times 10^{-4}\ mM$, the percentage loss of SNc somas and terminals were similar which was in the range of $50 - 60\%$, and these are the extracellular LDOPA concentrations that were observed in previous studies (Cullen and Wong-Lin, 2015; Khor and Hsu, 2007; Reed et al., 2012).

### 3.6. Insights into the Mechanism of LDOPA-Induced Toxicity

In order to test the hypothesis of LDOPA-induced toxicity, we have administered ranges of external LDOPA in the proposed model when the percentage loss of somas or terminals crosses 25% due to energy deficiency. When external LDOPA concentration ($36 \times 10^{-4}\ mM$) administered was near basal value, it was observed that the percentage loss of SNc somas and terminals was not altered much. When external LDOPA concentration administered was in the range from $36 \times 10^{-4}\ mM$ to $36 \times 10^{-3}\ mM$, it was observed that the percentage loss of SNc somas and terminals was decreasing. When external LDOPA concentration administered was above $36 \times 10^{-3}\ mM$, it was observed that the percentage loss of SNc somas was similar, but the percentage loss of SNc terminals was higher when external LDOPA concentration was near basal value (Figure 7B).

From simulation results, it was observed that LDOPA indeed induced toxicity in SNc cells at higher concentrations, which might be due to excitotoxicity or oxidative stress or both. To evade LDOPA toxicity in all stages of LDOPA therapy in the case of PD, we need to understand the mechanism behind the toxicity. In order to do so, we co-administered two different drugs along with LDOPA, namely SP antagonists and glutathione, which led to target overexcitation in SNc somas (by reducing SP-mediated excitatory inputs to SNc) and ROS build up in SNc terminals (by scavenging ROS) respectively. When SP antagonists are co-administered, it was observed that the percentage loss of SNc somas was decreasing with increasing inhibition of SP transmission (Figure 7C, blue bar). However, there was a change in the percentage loss of SNc terminals (Figure 7C, orange bar). When glutathione was co-administered, it was observed that the percentage loss of SNc somas and terminals was decreasing with increasing glutathione concentration (Figure 7D).

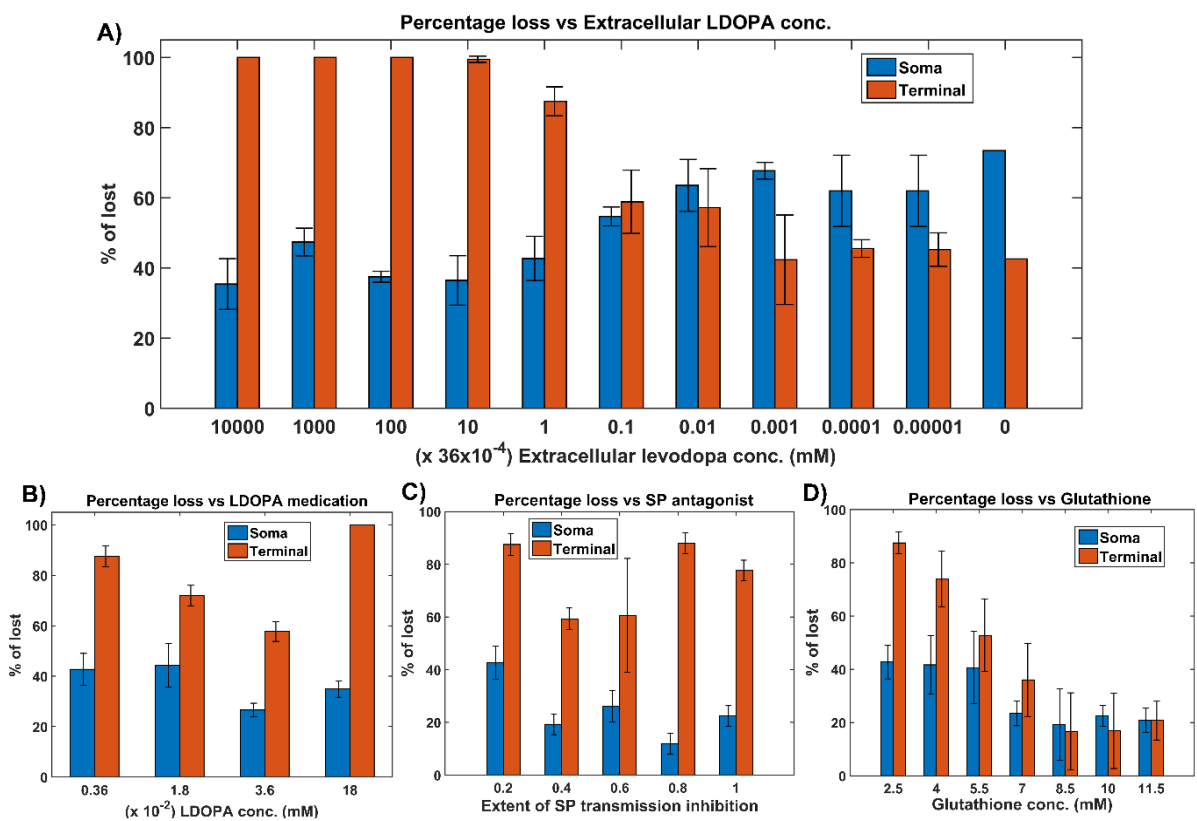

*Figure 7: Model response to different therapeutics under energy deficiency. **(A)** Percentage loss for various extracellular LDOPA concentrations, **(B)** Percentage loss during LDOPA therapy, **(C)** Percentage loss during SP antagonist therapy, **(D)** Percentage loss during glutathione therapy. All the therapeutic interventions were initiated at 25% soma or terminal loss. conc, concentration; LDOPA, levodopa; SP, substance P; mM, millimolar.*

# 4. DISCUSSION

## 4.1. LIT model

The goal of this computational study is to develop a model of SNc-striatum, which helps us in understanding LDOPA-induced toxicity in SNc neurons under energy deficiency conditions. From both homogeneous and heterogeneous energy deficiency results, it suggests that SNc (axonal) terminals are more vulnerable to energy imbalance when compared to SNc cell bodies (somas) which was observed experimentally, where injury is initiated at axonal terminals (Burré et al., 2010; Cheng et al., 2010; Giguère et al., 2019; Wong et al., 2019). The higher positive currents from STN might lead to excitotoxic loss of SNc somas (Figure 5A, blue bar), and increased ROS production might lead to increased SNc terminal loss (Figure 5A, orange bar). DA transporters, which play a crucial role in DA neurotransmission, were depleted more in axonal terminals compared to cell bodies in early PD (Fazio et al., 2018). From these studies, it can be postulated that pathogenesis starts at axonal terminals, which are more vulnerable to energy deficiencies and therefore are ideal sites for developing novel disease-modifying therapeutics.

When lower concentrations of extracellular LDOPA were available, the loss of SNc somas was more when compared to SNc terminals (Figure 7A). This might be due to lower extracellular DA levels as a result of lower extracellular LDOPA concentrations and lower vesicular DA levels (due to reduced packing of DA into vesicles as a result of lower energy levels) causing disinhibition of SNc somas (as result of lesser cortical excitation of MSNs) which are already in a low energy state. Due to disinhibition and energy deficiency, SNc somas might become overactive, which leads to calcium build-up, resulting in excitotoxic loss of SNc somas (Albin and Greenamyre, 1992; Muddapu et al., 2019). When higher concentrations of extracellular LDOPA were available, the loss of SNc terminals was more when compared to SNc somas (Figure 7A). This might be due to higher cytoplasmic DA levels as a result of higher extracellular LDOPA concentrations, lower vesicular packing of DA (due to lower energy levels) and LDOPA-induced stimulation of DA metabolism (Mosharov et al., 2009) resulting in DA-mediated oxidative stress in the SNc terminals (Farooqui, 2012; Morrison et al., 2012). Due to higher DA levels and energy deficiency, DA in SNc terminals causes oxidative stress, which leads to SNc terminal loss. At higher concentrations of extracellular LDOPA, loss of SNc somas was lower compared to lower concentrations of extracellular LDOPA as a result of the restoration of inhibitory tone from MSNs due to higher extracellular DA concentrations. The extracellular LDOPA concentration ranging from $36 \times 10^{-7}\ mM$ to $36 \times 10^{-4}\ mM$ was considered as basal extracellular LDOPA concentrations in the proposed LIT model. At these values, the percentage loss of SNc somas and terminals was similar, which was observed in previous studies (Cullen and Wong-Lin, 2015; Khor and Hsu, 2007; Reed et al., 2012). Our model was able to show the significance of basal extracellular LDOPA concentrations, which needed to be maintained for normal functioning.

When external LDOPA concentration administered was in the range from $36 \times 10^{-4}\ mM$ to $36 \times 10^{-3}\ mM$, it was observed that the percentage loss of SNc somas and terminals was decreasing, which suggests the therapeutic benefits of LDOPA therapy in altering or halting the progression of the SNc cell loss. When external LDOPA concentration administered rated was above this range, the neuroprotective effect of LDOPA therapy diminished. From LDOPA+SP antagonist co-administration, it was observed that inhibiting excitotoxicity in SNc somas did not decrease SNc terminal loss, which suggests that excitotoxicity in SNc somas does not contribute to oxidative stress in SNc terminals in LDOPA-induced toxicity. From LDOPA+glutathione co-administration, it was observed that inhibiting oxidative stress in SNc terminals did decrease the loss of SNc somas, which suggests that oxidative stress in SNc terminals does contribute to excitotoxicity in SNc somas in LDOPA-induced toxicity. From these results, we can suggest that adjunct therapies such as antioxidants (Betharia et al., 2019; Borah and Mohanakumar, 2010; Carvey et al., 1997; Deng et al., 2020; Nikolova et al., 2019; Pardo et al., 1993, 1995; Walkinshaw and Waters, 1995) and other potential therapies such as D2 agonists (Asanuma et al., 2003), Glycogen synthase kinase 3 inhibitors (Choi and Koh, 2018), calcium-binding protein drugs (Isaacs et al., 1997), etc. co-administrated along with LDOPA should be able to evade LDOPA toxicity in all stages of PD.

From the simulation results, it was observed that the LDOPA-induced toxicity in cell bodies and axonal terminals of SNc neurons was autoxidation-irrelevant and autoxidation-relevant, respectively. In the case of cell bodies, excess DA in striatum due to LDOPA therapy stimulates glutamatergic cortical inputs to MSNs, which leads to overexcitation of MSNs. The overexcited MSNs co-release SP along with GABA onto SNc neurons. SP modulates SNc glutamatergic inputs in such a way that it overexcites SNc neurons resulting in excitotoxic neuronal loss in SNc. However, in the case of axonal terminals, excess DA in terminals due to LDOPA therapy leads to autooxidation of DA. The autoxidation of DA results in the production of free radicals, which lead to oxidative stress in SNc axonal terminals resulting in axonal synaptic pruning of SNc neurons. The study suggests that LDOPA-induced toxicity occurs by two mechanisms: DA-mediated oxidative stress in axonal terminals of SNc neurons and by exacerbating STN-mediated overexcitation in cell bodies of SNc neurons.

## 4.2. SNc positive feedback loops – Scope of vulnerability

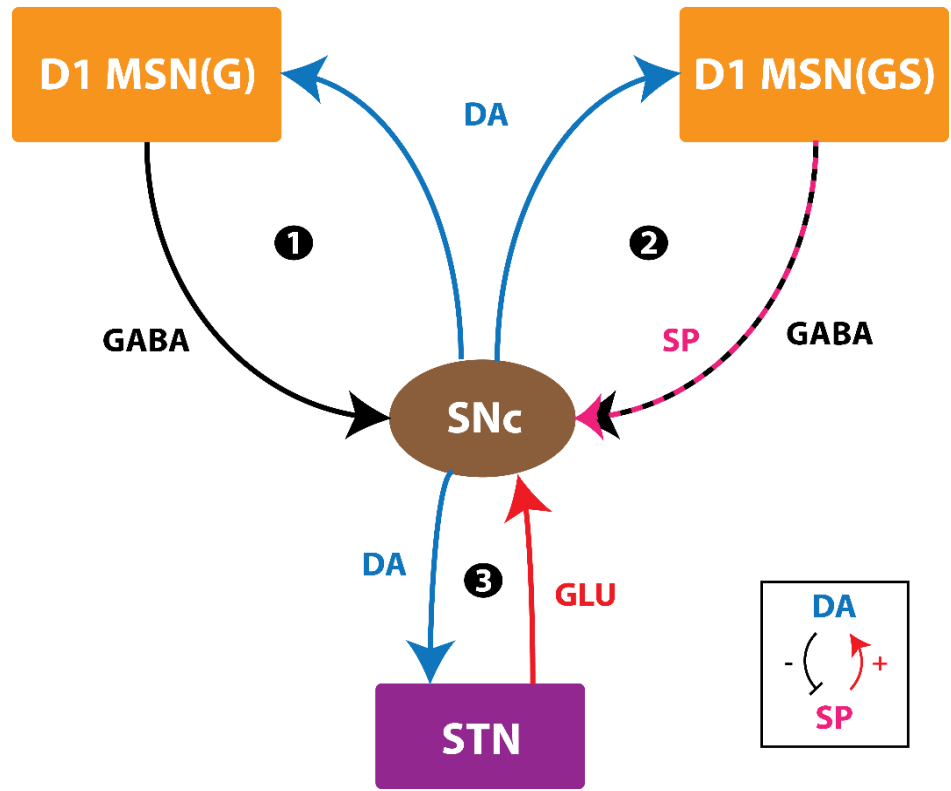

*Figure 8: Positive feedback loops of SNc. SNc, substantia nigra pars compacta; STN, subthalamic nucleus; D1 MSN(G), D1-type DA receptor-expressing medium spiny neuron (release GABA); D1 MSN(GS), D1-type DA receptor-expressing medium spiny neuron (co-release GABA and SP); DA, dopamine; SP, substance P; GLU, glutamate; GABA, gamma-aminobutyric acid. Inset, DA-SP feedback.*

In normal conditions, there is no SNc cell or terminal loss where SNc maintains the dopaminergic tone on its target regions such as STN, D1-MSN(G), and D1-MSN(GS). In the first loop (Figure 8), normal dopaminergic tone to D1-MSN(G) results in inhibition of SNc by GABAergic projections. In the second loop (Figure 8), normal dopaminergic tone to D1-MSN(GS) results in inhibition of SNc by GABA and lesser excitation of SNc by SP due to DA-SP feedback (Brimblecombe and Cragg, 2015; Thornton and Vink, 2015). In the third loop (Figure 8), normal dopaminergic tone to STN results in lesser excitation of SNc by glutamatergic projections (Hassani et al., 1997; Magill et al., 2001; Yang et al., 2016).

Under pathological conditions, there is SNc cell or terminal loss where SNc fails to maintain the dopaminergic tone in its target regions such as STN, D1-MSN(G), and D1-MSN(GS). In the first loop (Figure 8), DA deficiency in the striatum causes lesser excitation of D1-MSN(G) by cortex, which, by feedback, results in disinhibition of SNc. In other words, initial DA deficiency due to SNc cell loss causes lesser excitation of D1-MSN(G), which in turn disinhibits SNc resulting in further SNc cell loss due to excitotoxicity, which acts as positive feedback. In the second loop, DA deficiency in the striatum causes lesser excitation of D1-MSN(GS) by cortex, which results in disinhibition (through GABA) and further excitation (through SP; due to low DA, the effect of SP gets enhanced) of SNc. Thus, the disinhibition of SNc happens in a manner similar to the first loop; however, the overexcitation of SNc happens due to the DA-SP feedback mechanism, which also acts as positive feedback. In the third loop, DA deficiency causes overexcitation of STN, which results in overactivation of SNc. In other words, initial DA deficiency due to SNc cell loss causes overexcitation of STN, which in turn overexcites SNc, by a positive feedback mechanism, resulting in further SNc cell loss due to excitotoxicity.

In medication conditions, LDOPA is administrated where dopaminergic tone to SNc target regions (STN, D1-MSN(G), D1-MSN(GS)) is restored. If the administrated LDOPA dosage goes beyond a certain threshold, overexcited D1-MSN(GS) through the DA-SP feedback mechanism makes SNc hyperactive, which in turn results in SNc cell loss due to excitotoxicity. Along with SNc cell body loss, SNc terminals also undergo degeneration due to excess DA causing oxidative stress. To summarize, LDOPA-induced toxicity in SNc doesn't occur when LDOPA dosage is below the threshold, which results in the survival of remaining SNc cells. However, if, LDOPA dosage goes beyond a threshold, from that point onwards, the aforementioned runaway effect kicks in, leading to a progressive and irrevocable cell loss in SNc. Thus, it is evident that LDOPA might be toxic to SNc neurons under high dosage, which triggers a runaway effect resulting in uncontrollable SNc cell loss.

### 4.3. Limitations and Future Directions

Though the proposed model captures the exciting results of LDOPA-induced toxicity, it is not without limitations. In the proposed model, for example, the serotonergic system was not considered, which takes up LDOPA and contributes to striatal DA levels (Stansley and Yamamoto, 2015; Svenningsson et al., 2015) that contribute to LDOPA-induced dyskinesias (Carta et al., 2008; Carta and Tronci, 2014). Similarly, interneurons in the striatum were also not considered for simplifying the model.

In the proposed model, the ischemic condition was implemented by lowering glucose and oxygen levels, which can be extended by adding a blood vessel module (Cloutier et al., 2009) and varying cerebral blood flow to simulate ischemia condition more realistically. In the proposed model, stress was monitored in SNc neurons alone, which can be extended to other neuronal types in the model by monitoring stress levels, where intracellular calcium build-up can be a stress indicator (Bano and Ankarcrona, 2018). In order to do so, all neuronal types should be modelled as conductance-based models where calcium dynamics should be included. From our studies, it is shown that LDOPA dosage plays an important role in the progression of the disease. The proposed model will be integrated with a behavioural model of cortico-basal ganglia circuitry (Muralidharan et al., 2018) to show the effect of LDOPA-induced toxicity at the behavioural level and optimize the LDOPA dosage so as to achieve maximum effect with a minimal dosage of the drug.

We suggest some experimental approaches to validate some of the predictions from our modelling study. Under induced progressive energy deficiency conditions in animal models (Puginier et al., 2019), LDOPA administration at moderate levels can also be toxic, which needs to be studied by measuring metabolites of DA autoxidation process. In order to study the effects of LDOPA-induced toxicity in SNc somas in midbrain and SNc terminals in the striatum, similar toxin-induced animal models can be used, where oxidative stress in terminals can be examined by monitoring the levels of free radicals and excitotoxicity in somas can be examined by monitoring calcium levels (Wong et al., 2019). By co-administering antioxidants along with LDOPA in toxin-induced animal models (Betharia et al., 2019; Borah and Mohanakumar, 2010; Carvey et al., 1997; Nikolova et al., 2019; Pardo et al., 1993, 1995; Walkinshaw and Waters, 1995), the progression of SNc soma and terminal loss can be altered along with prolonging the 'honeymoon period' of LDOPA therapy (Erro et al., 2016; Holford and Nutt, 2008; Stocchi et al., 2010).

## 5. CONCLUSIONS

In conclusion, we believe that the proposed model provides significant insights in understanding the mechanisms behind LDOPA-induced toxicity under energy deficiency conditions. From simulation results, it was shown that SNc terminals are more vulnerable to energy imbalances when compared to SNc somas. The study suggests that LDOPA-induced toxicity occurs differently in SNc somas and terminals: in the case of SNc somas, it might be due to excitotoxicity caused by STN, and in case of SNc terminals, it might be due to oxidative stress caused by dopamine autoxidation. From adjunct therapies, it was clear that co-administering antioxidants, along with LDOPA, can be neuroprotective. By the aforementioned modelling efforts in addition to some earlier ones (Muddapu et al., 2019), we are trying to understand the root cause of PD neurodegeneration as energy deficiency occurring at different neural hierarchies: subcellular, cellular and network levels.

**AUTHOR CONTRIBUTIONS**

VRM - Conceptualization; Model development; Data curation; Formal analysis; Investigation; Methodology; Manuscript writing; VSC - Conceptualization; Model development; Data curation; Formal analysis; Investigation; Methodology; Manuscript writing; Supervision; KV - Conceptualization; Model development; KR - Conceptualization; Model development;

**Table 1:** Different population sizes in the proposed LIT model.

| Network type | Size (# of nodes) |
|:---:|:---:|
| SNc (soma) | $8 \times 8$ (64) |
| SNc (terminal) | $32 \times 32$ (1024) |
| D1-MSN (G) | $32 \times 32$ (1024) |
| D1-MSN (GS) | $32 \times 32$ (1024) |
| STN | $32 \times 32$ (1024) |
| GPe | $32 \times 32$ (1024) |
| CTX | $32 \times 32$ (1024) |

**Table 1:** Parameter values used in the proposed model of LIT.

| Parameter(s) | STN | GPe | CTX | MSN |
|---|---|---|---|---|
| Izhikevich parameters | | | | |
| $a\ (ms^{-1})$, | $a = 0.005$, | $a = 0.1$, | $a = 0.03$, | $a = 0.01$, |
| $b\ (pA.mV^{-1})$, | $b = 0.265$, | $b = 0.2$, | $b = -2$, | $b = -20$, |
| $c\ (mV)$, | $c = -65$, | $c = -65$, | $c = -50$, | $c = -55$, |
| $d\ (pA)$ | $d = 1.5$ | $d = 2$ | $d = 100$ | $d = 91$ |
| External current $(I^x)$ | $3\ pA$ | $4.25\ pA$ | $100\ pA$ | $0\ pA$ |
| Maximum peak of voltage $(v_{peak}^x)$ | $30\ mV$ | $30\ mV$ | $40\ mV$ | $35\ mV$ |
| Membrane capacitance $(C^x)$ | $1\ \mu F$ | $1\ \mu F$ | $100\ \mu F$ | $15.2\ pF$ |
| Resting potential $(v_r^x)$ | - | - | - | $-80\ mV$ |
| Threshold potential $(v_t^x)$ | - | - | - | $-29.7\ mV$ |
| Membrane constant $(k^x)$ | - | - | - | $1\ pA.mV^{-1}$ |
| Number of laterals $(nlat^x)$ | 11 | 15 | - | - |
| Radius of Gaussian laterals $(R^x)$ | 1.4 | 1.6 | - | - |
| Synaptic strength within laterals $(A^x)$ | 1.3 | 0.1 | - | - |
| Time decay constant for AMPA $(\tau_{AMPA})$ | $6\ ms$ | $6\ ms$ | $6\ ms$ | $6\ ms$ |
| Time decay constant for NMDA $(\tau_{NMDA})$ | $160\ ms$ | $160\ ms$ | $160\ ms$ | $160\ ms$ |
| Time decay constant for GABA $(\tau_{GABA})$ | $4\ ms$ | $4\ ms$ | $4\ ms$ | $4\ ms$ |
| Synaptic potential of AMPA receptor $(E_{AMPA})$ | $0\ mV$ | $0\ mV$ | $0\ mV$ | $0\ mV$ |
| Synaptic potential of NMDA receptor $(E_{NMDA})$ | $0\ mV$ | $0\ mV$ | $0\ mV$ | $0\ mV$ |
| Synaptic potential of GABA receptor $(E_{GABA})$ | $-60\ mV$ | $-60\ mV$ | $-60\ mV$ | $-60\ mV$ |
| Concentration of Magnesium $(Mg^{2+})$ | $1\ mM$ | $1\ mM$ | $1\ mM$ | $1\ mM$ |

**Table 1:** Connectivity patterns in the proposed LIT model.

| From – to | Pattern (signal) |
|---|---|
| SNc (soma) – SNc (terminal) | 1 to 16 (Calcium) |
| SNc (terminal) – D1-MSN (GS) | 20 to 1 (Dopamine) |
| SNc (terminal) – D1-MSN (G) | 20 to 1 (Dopamine) |
| D1-MSN (GS) – D1-MSN (G) | 1 to 1 (GABA & SP) |
| D1-MSN (G) – D1-MSN (GS) | 1 to 1 (GABA) |
| D1-MSN (GS) – SNc (soma) | 200 to 1 (GABA & SP) |
| D1-MSN (G) – SNc (soma) | 200 to 1 (GABA) |
| STN – GPe | 1 to 1 (Glutamate) |
| GPe – STN | 1 to 1 (GABA) |
| STN – SNc (soma) | 16 to 1 (Glutamate) |
| CTX – D1-MSN (GS) | 1 to 1 (Glutamate) |
| CTX – D1-MSN (G) | 1 to 1 (Glutamate) |
| STN – STN | Gaussian neighborhoods (Glutamate) |
| GPe – GPe | Gaussian neighborhoods (GABA) |
| SNc – SNc | Gaussian neighborhoods (GABA) |

**Table 1:** Parameter values used in the proposed model of LIT.

| Parameter | Value | Parameter | Value |
|---|---|---|---|
| Number of laterals ($nlat^x$) | 5 | $\theta_g$ | $20\ mV$ |
| Radius of Gaussian laterals ($R^x$) | 1.6 | $\theta_g^H$ | $-57\ mV$ |
| Synaptic strength within laterals ($A^x$) | 0.1 | $\sigma_g^H$ | $2\ mV$ |
| Synaptic conductance ($W_{x \to y}$) | 0.01 | $\alpha$ | $2\ ms^{-1}$ |
| Synaptic potential of GABA receptor ($E_{GABA}$) | $63.45\ mV$ | $\beta$ | $0.08\ ms^{-1}$ |
| $s_{max}^{STN}$ | 1.3 | $cd_{stn}$ | 4.87 |
| $s_{min}^{GPe}$ | 0.1 | $cd_{gpe}$ | 7 |
| $s_{min}^{SNc}$ | $1 \times 10^{-6}$ | $cd_{snc}$ | 4.6055 |
| $cd2$ | 0.1 | $w_{sp}$ | 5000 |
| $K^{MSN}$ | 0.0289 | $L^{MSN}$ | 0.331 |
| $\alpha_{DA}^{D1-MSN(G)}$ | 1 | $\alpha_{DA}^{D1-MSN(GS)}$ | 2 |
| $w_{GPe \to GPe}$ | 1 | $w_{SNc \to SNc}$ | 0.01 |
| $w_{STN \to GPe}$ | 1 | $w_{GPe \to STN}$ | 20 |
| $w_{STN \to STN}$ | 1.3 | $w_{STN \to SNc}$ | 0.3 |
| $w_{D1-MSN\ (G) \to SNc}$ | 0.5 | $w_{D1-MSN\ (GS) \to SNc}$ | 0.5 |
| $w_{D1-MSN\ (G) \to D1-MSN\ (GS)}$ | 500 | $w_{CTX \to D1-MSN\ (GS)}$ | 100 |
| $w_{CTX \to D1-MSN\ (G)}$ | 100 | $\tau_d^{sp}$ | $40\ ms$ |
| $\tau_f^{sp}$ | $200\ ms$ | $\tau_r^{sp}$ | $10\ ms$ |
| $\beta_{sp}$ | 0.47 | $\lambda_{sp}$ | 5.5 |
| $b_{sp}$ | 2.5 | $F_{STN \rightarrow SNc}$ | $1 \times 10^{-5}$ |
| $F_{D1-MSN\,(G) \rightarrow SNc}$ | $4.15 \times 10^{-6}$ | $F_{D1-MSN\,(GS) \rightarrow SNc}$ | $4.15 \times 10^{-6}$ |
| $V_{trans}^{max}$ | $5.11 \times 10^{-7}\ mM.ms^{-1}$ | $K_m^{LDOPA_s}$ | $0.032\ mM$ |
| $[TYR_s]$ | $0.063\ mM$ | $K_a^{TYR_s}$ | $0.064\ mM$ |
| $[TRP_s]$ | $0.082\ mM$ | $K_a^{TRP_s}$ | $0.015\ mM$ |
| $P_{soma}^{snc}$ | 64 | $P_{terminal}^{snc}$ | 1024 |
| $ER_{thres}$ | $2.15 \times 10^{-3}\ mM$ | $ROS_{thres}$ | $0.0147\ mM$ |
| $MT_{thres}$ | $0.0215\ mM$ | | |